\documentstyle[preprint,aps]{revtex}

\draft

\title{Bolometric detection of mechanical bending waves in suspended carbon nanotubes}

\author{B. Reulet$^1$, A.Yu. Kasumov$^{1,2}$, M. Kociak$^1$,
R. Deblock$^1$, I.I. Khodos$^2$,\\ Yu.B. Gorbatov$^2$,
V.T. Volkov$^2$, C.Journet$^3$ and H. Bouchiat$^1$}

\address{$^1$Laboratoire de Physique des Solides, Associ\'e au CNRS,
B\^at 510, Universit\'e Paris--Sud, 91405, Orsay, France.$
^2$Institute of Microelectronics Technology and High Purity Materials,
Russian Academy of Sciences, Chernogolovka 142432 Moscow Region,
Russia.$^3$ Groupe de Dynamique des Phases Condens\'ees, Universit\'e
Montpellier II 34095 Montpellier France. }

\begin{document}

\maketitle 

\begin{abstract}

We show that it is possible to detect mechanical bending modes on
 $1\mu m$ long ropes of
single walled-carbon nanotubes suspended between 2 metallic contacts.
This is done by measuring either their dc resistance in a
region of strong temperature dependence 
(in the vicinity of superconducting or metal-insulator transition),
or their critical current.
The vibrations are excited by a radio-frequency electric field 
produced by an antenna located in the vicinity of the sample. 
We analyze the mechanism of detection of the mechanical resonances in
terms of heating and phase breaking effects.
\end{abstract}

\newpage

Single walled Carbon nanotubes (SWNT) are molecular wires which
combine both remarkable electronic and mechanical properties.
Depending of its diameter and helicity a SWNT can be either
semiconducting or metallic with only two conducting modes at the Fermi
energy \cite{dressel,dekker}. The Young modulus of a rope constituted of small number of
aligned SWNT has been estimated from the study of vibrations by electron microscopy
\cite{treacy,yao} and more recently by  force
microscopy experiments \cite{salvetat} to be as high as 1TPa. This result is in
agreement with theoretical calculations \cite{liu}.  In the following we show that it is
possible to excite and detect stationary mechanical bending modes of a
SWNT rope which is suspended between metallic contacts across a
slit. The excitation is produced by applying a radio-frequency electromagnetic field
to the tube. The energy dissipated in the sample at resonance is
detected by measuring the dc resistance of the tube in a range of
strong temperature dependence and the critical current for the superconducting ropes.
 
We use as starting material carbon nanotubes prepared by the
electrical arc method. When cobalt is used as a catalyst
\cite{journet}, this method produces single wall tubes whose diameters
are all of the order of 1nm. In most cases these tubes are assembled
into ropes containing typically 100 nearly ordered parallel tubes.
  
Isolation of an individual rope and its connection to electrical
contacts were performed according to the following procedure: a
focused laser beam releases a nanotube from a target onto the sample
with a $Si_3N_4$ membrane coated with a metal film. A submicron width
slit about $100 \mu m$ in length has previously been cut in the
membrane using a focussed ion beam and has been disposed under the
tube on the target using an optical microscope.  After deposition, the
nanotube connects the edges of the slit and shorts the electric
circuit whose resistance was over $1 G\Omega$ before the nanotube was
deposited.  This procedure has been shown to produce low resistivity
ohmic contacts between the tube and the metal of the electrodes which
is melt during the process \cite{kasu2,kasu3}.  A transmission electron microscopy image
of such a sample is shown in fig.1A.  Depending on the value
of the resistance of the rope at room temperature, different temperature
dependences are observed at low temperature, going from a thermally
activated behavior at low temperature for resistive samples to a quasi
metallic behavior where the resistance is temperature independent.
When low resistive ropes are connected to superconducting contacts
they become superconducting due to  proximity effect with zero resistance
below the transition temperature of the contacts (see fig.1B).  

In the following we will restrict our discussion on the  two samples $RO_1$ and $RO_2$
mounted on Au/Re contacts  superconducting at 1.2K. $R0_1$ is $1.7
\mu m$ long and contains approximately $200$ SWNT , the room
temperature value of its resistance is equal to $65\Omega $ which is
only twice the expected value assuming that each tube is a two-channel ballistic wire. 
$R0_1$ becomes superconducting  courtesy of the proximity effect, below  the transition
temperature of Au/Re \cite{kasu3}.  On the other hand $R0_2$ ($0.6\mu m$ long and
constituted of $100$ SWNT), is much more resistive ($10 k\Omega$ at
room temperature) and exhibits an exponential divergence of its
resistance at low temperature (see fig. 1B).
The dc transport properties of these ropes are
strongly affected by the presence of a radio-frequency
electromagnetic field produced by an antenna located in the vicinity
of the sample.
  
The dc voltage drop across the superconducting rope $R0_1$ is depicted on fig.2 as
a function of the rf electric field frequency which amplitude is estimated to
be of the order of $E_{rf}=30mV/cm$. The sample is fed by a dc current
of $2\mu A$ (below the
critical value $i_c=2.7 \mu A$ \cite{kasu3}). One can
clearly identify a succession of resonance peaks which are harmonics
of $f_1= 330Mhz$ . When the rf power is not too high these can be
described by a lorentzian line-shape with a quality factor of the order
of 1000 at 100mK and decreasing with temperature like 1/T in the
superconducting regime (see inset of fig. 2).
These resonances are not observed in
the normal state where the resistance is temperature
independent. Similar type of resonances are also observed on
semi-conducting tubes. An example is shown on fig.3B with the semiconducting sample
$R0_2$: the dc resistance function of the RF frequency presents
a negative peak at 860MHz. In both cases,
we interpret these resonances as stationary mechanical bending
modes of the rope. The fundamental mode of a vibrating rod attached on
both ends is indeed given by \cite{landau}: 

\begin{equation}
f_1=22.4 \frac{R}{2L^2}\sqrt{\frac{Y}{\rho}}
\end{equation}
where $Y$ is the Young modulus estimated
to $10^{12} Pa$  \cite{salvetat,liu}, $\rho=1.3 kg/m^3$ is the mass per
unit volume, $L$ and $R$ the length and radius of the rod
($L=1.7\mu m$, $R=7nm$ for $RO_1$ and $L=0.6\mu m$, $R=2.5nm$ for $RO_2$).
This yields to $f_1=280MHz$ for $RO_1$ and $f_1=1.2GHz$ for $R0_2$.
Note that for $R0_1$ the resonance frequencies
experimentally observed are of the order of $n f_1$ with
$n>1$. The fundamental frequency is not observed, which may be
due to  very small coupling between the antenna and the tube  in this frequency range.
On the other hand only the fundamental
mode could be observed for $R0_2$.  The mechanical nature of the
resonances has been confirmed for $R0_1$ by injecting a small amount
of nitrogen gas in the vacuum can of the dilution fridge. Adsorption
of nitrogen atoms on the tube resulted in a small drift of the
resonances down to lower frequency, which is indeed expected when the
mass of the tube is slightly increased. A similar frequency shift of the
resonance could also be detected on $R0_2$ at T=4.2K when the sample
is immersed from  helium gas into liquid helium (see fig.3). This measured relative
shift ($\delta f/f=-0.012$) is smaller but of the order of the calculated
value -0.03 taking an increase of the effective density of the SWNT equal
to the density of liquid helium. It is possible that the SWNT already
initially contains a certain amount of adsorbed helium atoms.
 
An important issue for the understanding of these observations is the
mechanism of conversion of the electromagnetic field into a transverse
ultrasonic wave through the tube. We could not detect any observable change in the
resonance spectrum when applying a dc voltage on the antenna superimposed on the
ac excitation . So, we suggest that the presence of electrostatic charges
on the tube and the Coulomb force
produced  by the radio-frequency electric field on these charges are the main
mechanism of excitation of mechanical vibrations \cite{lorentz}. 
The existence of charge depletion in a carbon nanotube in
close contact with a noble metal such as Au or Pt has been shown to
arise from the difference of electronic work functions between the
SWNT and the metal by Dekker et al. \cite{dekstm}. From this work, the
resulting uncompensated charges on a SWNT can be estimated to be of
the order of $q= 100e$ where $e$ is the electron charge.
It is possible that these charges are
not uniformly spread on the tube which could favor the excitation of
harmonics compared to the fundamental. A similar procedure for the excitation of
mechanical resonances has been investigated in cantilevered multiwalled nanotubes
under an electron microscope \cite{vibmic}.

As we discussed below the resonances can be detected by dc transport only in the
regions where the resistance of the tubes depends strongly on temperature. This
suggests that the mechanism underlying the detection of the signal is the heating
of the electrons, which is maximal at resonance. Nevertheless in that case the
line-shape of the resonances should depend strongly on the power of the rf field,
since as discussed below their quality factor strongly depends on temperature. This
is not what was measured: on the contrary, the quality factor remains constant as
the power is varied, until a critical value above which saturation effects appear, as
shown on fig. 4. This implies that the suspended tubes are probably poorly thermally 
connected to the contacts which are not heated by the rf power. Then
the central part of the sample is heated up to a temperature $T_{eff}$
whereas the ends remain at $T_0$.
Let us try now to estimate $T_{eff}$.
The force on the rope can be estimated from the knowledge of the rf
electric field $E_{rf}$ applied on the tubes: $F=(1/4\pi\epsilon_0)N q E_{rf}$.
This force yields to a vibration of amplitude $\delta x$ given by
$\delta x= Q F L^3/\lambda R^4 Y$
where $Q$ is the quality factor of the resonance and $\lambda$ a numerical
coefficient of the order of 300. The power $P_{diss}$ dissipated at a resonance
of frequency $f$ reads $P_{diss}= F\delta x 2\pi f/Q$.
This power is transferred to the electrons via
electron-phonon collisions in the tube which results in increasing
the temperature in the center of the sample. The thermal impedance
$Z_{th}$ between the tube and the superconducting contacts 
is dominated by the value of the superconducting part of the rope at
$T_0$ and be estimated as follows \cite{tinkham}:

\begin{equation}
Z_{th}= exp(\Delta/T_0) (T_0/\Delta)^2/N k_1(T_0) \sim 4 10^{17}W mK^{-1}
\end{equation}
where $N$ is the number of SWNT in the rope,
$ k_1(T)=k_B^2T/h$  is the thermal conductance of
a ballistic 1D wire, $\Delta$ is the superconducting gap for the
bilayer $Au/Re$ estimated to be of the order of $2K$ and $T_0=100mK$ the
temperature of the contacts.
From that value the effective temperature of the tube is
obtained from the relation: $T_{eff}-T_0=Z_{th}P_{diss}$ which gives
$T_{eff}\sim0.5K$, to be compared with the range of temperature where
the resistance of $RO_1$ increases from $0$ to $50\Omega$, i.e. $0.8-2.3K$.
We have neglected the contribution of the phonons to the thermal conductivity
This is justified by the observation of a strong decreasing of $T_{eff}$
for a given rf power when applying a magnetic field of the order of
the critical field of the contacts. A similar calculation can be done
on the rope $R0_2$, but for this semiconducting sample one has to
take into account the reduced value of $k_1$ in the normal
state according to the Wiedemann-Franz law. This yields to an
effective temperature of $T_{eff}=5K$ (for $T_0=4.2K$ and $E_{rf}=1V/cm$).

Note however that all this 
analysis relies on the assumption that it is possible to define a
temperature of the electrons. This is only true when
the inelastic electron-electron scattering length is short
compared to the total length of the tube. It is also possible 
that the vibrations of the tube are at the origin of
dephasing processes, like electron-phonon scattering. This results in
a reduction of the phase coherence length of the electrons below
the length of the tube.
The existence of such a phase breaking mechanism  more
efficient at low rf power than the simple heating of the tube for the
destruction of proximity induced superconductivity  is corroborated by the
study of the critical current through $R0_1$ versus
excitation power. As a matter of fact we observed an exponential decay
of the critical current, whereas its temperature dependence
follows a BCS type of behavior \cite{kasu3}, very
flat below $T_c/2$ (see fig. 4A). The critical
current through an SNS junction is indeed known to vary like
%\begin{equation}
$I_c\propto exp(-L/L_\phi))$
%\end{equation}
where $L_\phi$ denotes the phase coherence length, related to
the phase coherence time $\tau_\phi$ by $L_\phi=v_F\tau_\phi$ for a ballistic junction.
Thus our result can be understood as a contribution of
the rf phonons to the inverse coherence time as $\tau_\phi^{-1}\propto P_{rf}$.
Note that the dephasing is not directly related to the rf electric field but
to the induced vibrations at resonance. The same rf power
applied out of resonance has indeed no effect on the critical current.

The mechanical resonances on $RO_1$ are detected by measuring
the critical current for very low excitation power.
For higher power the critical current vanishes whereas a finite resistance
shows up. As shown in fig. 4B, there
is a narrow range of rf power for which the resistance exhibits a lorentzian shape as
a function of frequency. At even higher power the shape is no longer lorentzian
but presents a plateau in its center.
This can be understood considering that the ends of the rope are no longer at $T_0$.
When they become normal their thermal conductivity
increases drastically, resulting in a much better thermalization of the center of
the sample, whose temperature does not increase anymore and even can decrease,
as observed on fig. 4B.

We have shown that suspended nanotube molecules work as a passive
radio set tuned to definite resonant frequencies which correspond to
the mechanical transverse eigen-modes of the tube.  The detection of
these modes is done courtesy of the very low thermal conductance of
the tubes which behave as extremely sensitive bolometers when their resistance is
temperature dependent.  Moreover we have shown for the superconducting sample that a
more effective mechanism is involved, implying phase breaking effects. The next
step would be to investigate the influence of these low energy phonons
on transport properties by systematic comparison of the proximity
effect on deposited and suspended tubes. It has indeed been shown
\cite{martin} that coupling with low energy phonons can turn repulsive
interactions in a Luttinger liquid into attractive ones and drive the
system towards a superconducting phase.

\begin{figure}
%\[     \epsfbox{file.eps}      \]
\caption{A. Transmission electronic micrograph  of a rope of
          single wall nanotubes, suspended across a slit between 2
          metallic pads. From these pictures it is possible to confirm
          that the metal of the contacts neither covers nor penetrates
          the tubes and to estimate the length and diameter of the
          samples. B. Temperature dependence of the resistance of different ropes
         mounted on Au/Re contacts measured for zero magnetic
field. The arrow indicates the transition of the Au/Re bilayer. The
sample $R0_1$ becomes superconducting below 1K. \label{fig1}}
\end{figure} 
        
\begin{figure}
%\[     \epsfbox{file.eps}      \]
\caption{A. Effect of a radio-frequency  electromagnetic radiation on the dc voltage
across the rope $R0_1$ when it is run through by a dc current below the critical
current . Inset: Evolution of the resonance line shapes of the 6th harmonics
on $R0_1$  with the temperature of the contacts.\label{fig2}}
\end{figure}

\begin{figure}
                %\epsfbox{file.eps}     \]
\caption{A. Effect of the deposition of solid nitrogen onto the rope $R0_1$
detected on the dc resistance versus frequency of the electromagnetic radiation.
B. Effect of the immersion in liquid helium of the rope $RO_2$ detected on the
dc resistance versus frequency of the electromagnetic radiation.
\label{fig3}}
\end{figure}

\begin{figure}
%\[     \epsfbox{file.eps}      \]
\caption{A. Critical current measured on the rope $RO_1$, on one hand
as a function of the rf power at resonance of the 6th harmonics (lower curve),
and on the other hand as a function of the temperature (upper curve).
B. Resistance of the rope $RO_1$ versus rf frequency near the 6th harmonics,
for different applied powers.
\label{fig4}}
\end{figure}


\begin{thebibliography}{99}
\bibitem{lorentz}We have checked that the amplitude of these resonances do not
depend on the current through the tube. The Lorentz force produced by the ac
magnetic component of the RF field acting  on the rope (run through by a dc
current of the order of $10nA$) is  negligible compared to the Coulomb
force discussed.
\end{thebibliography}
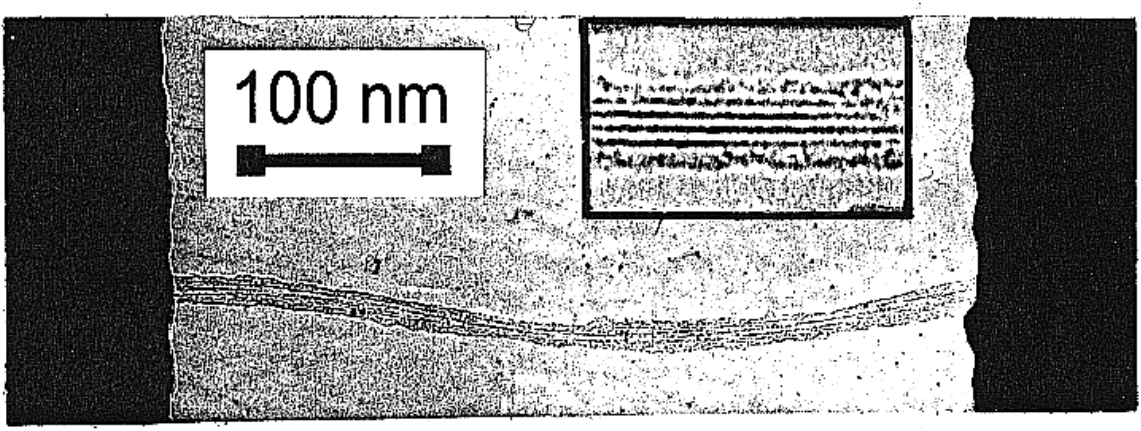
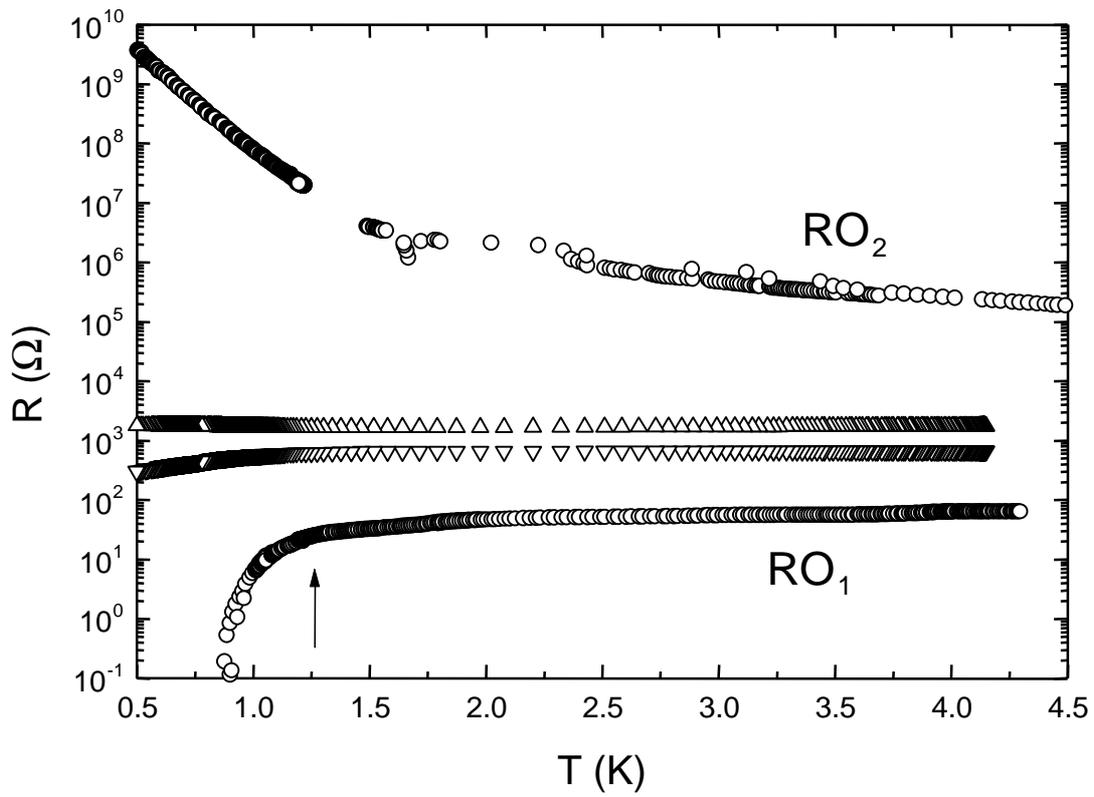
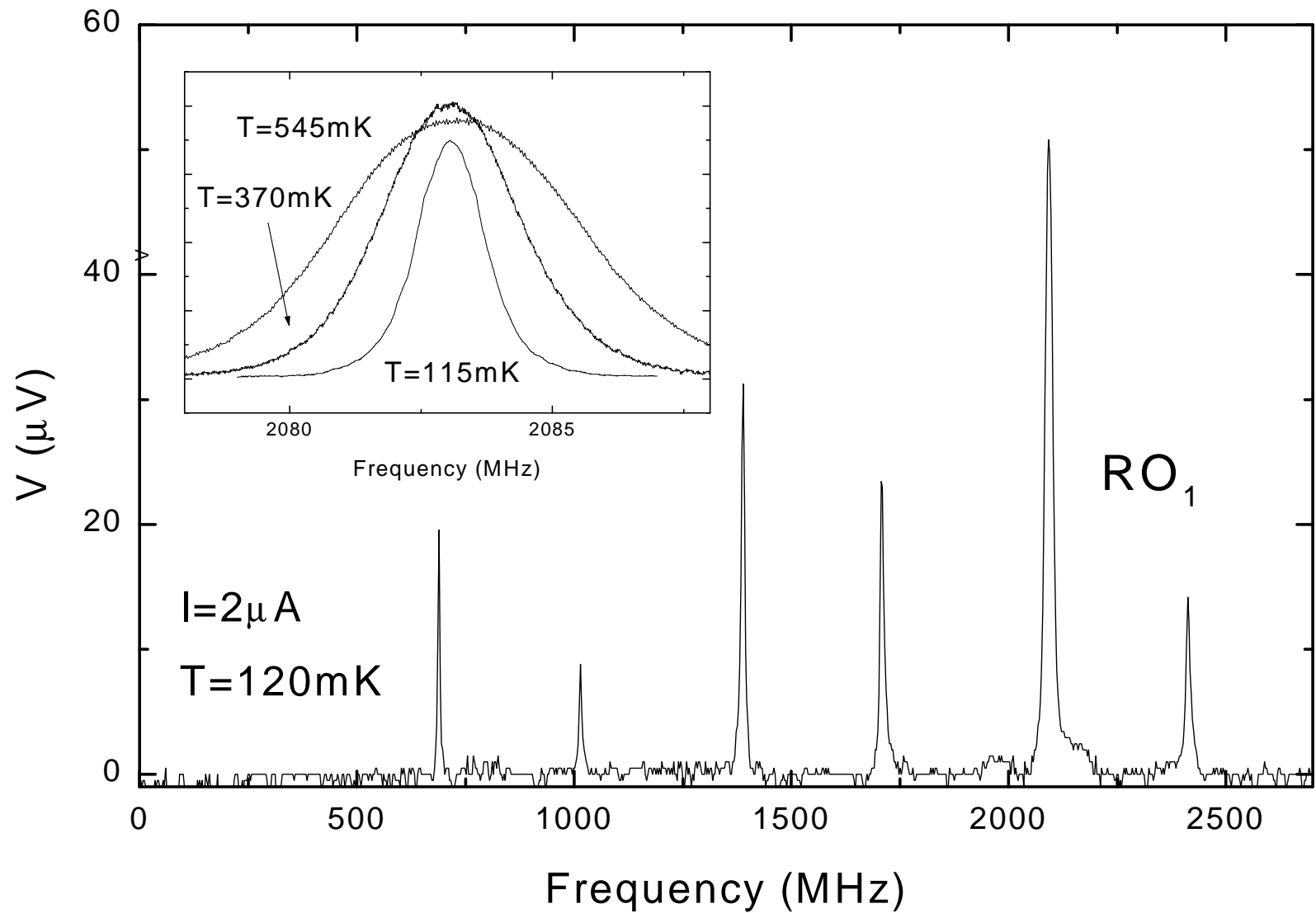
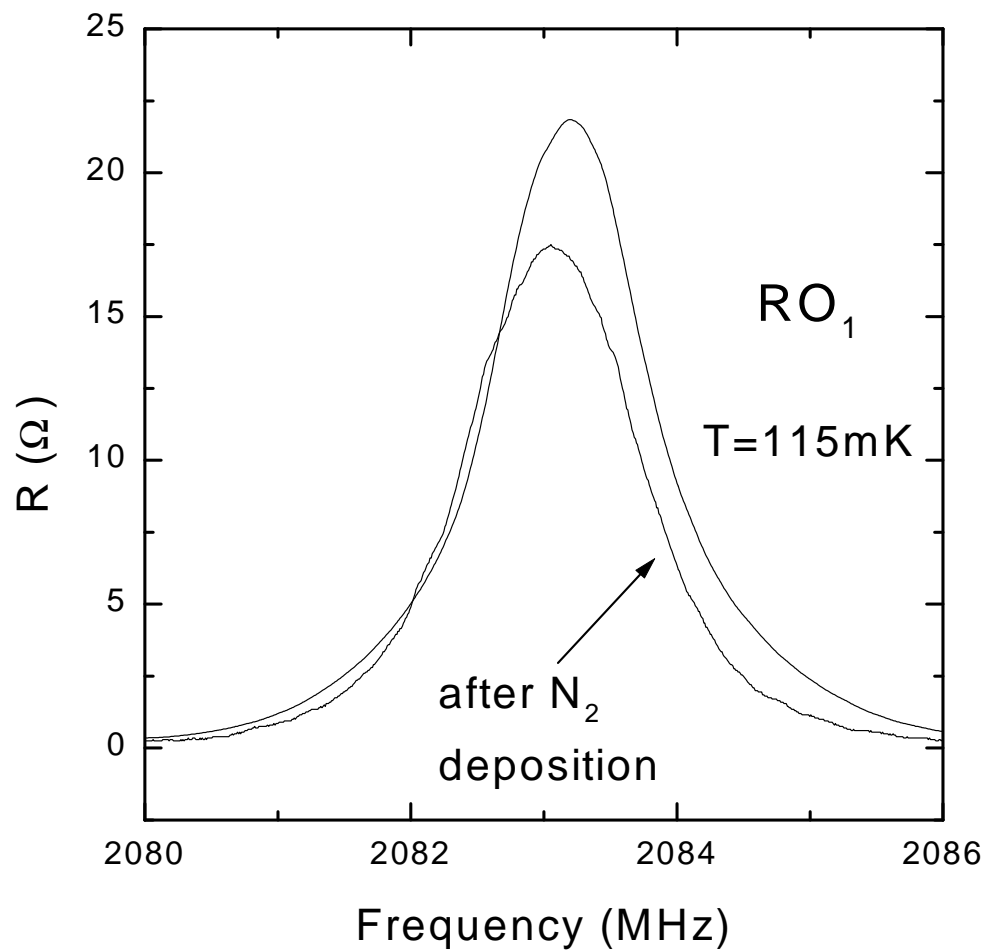
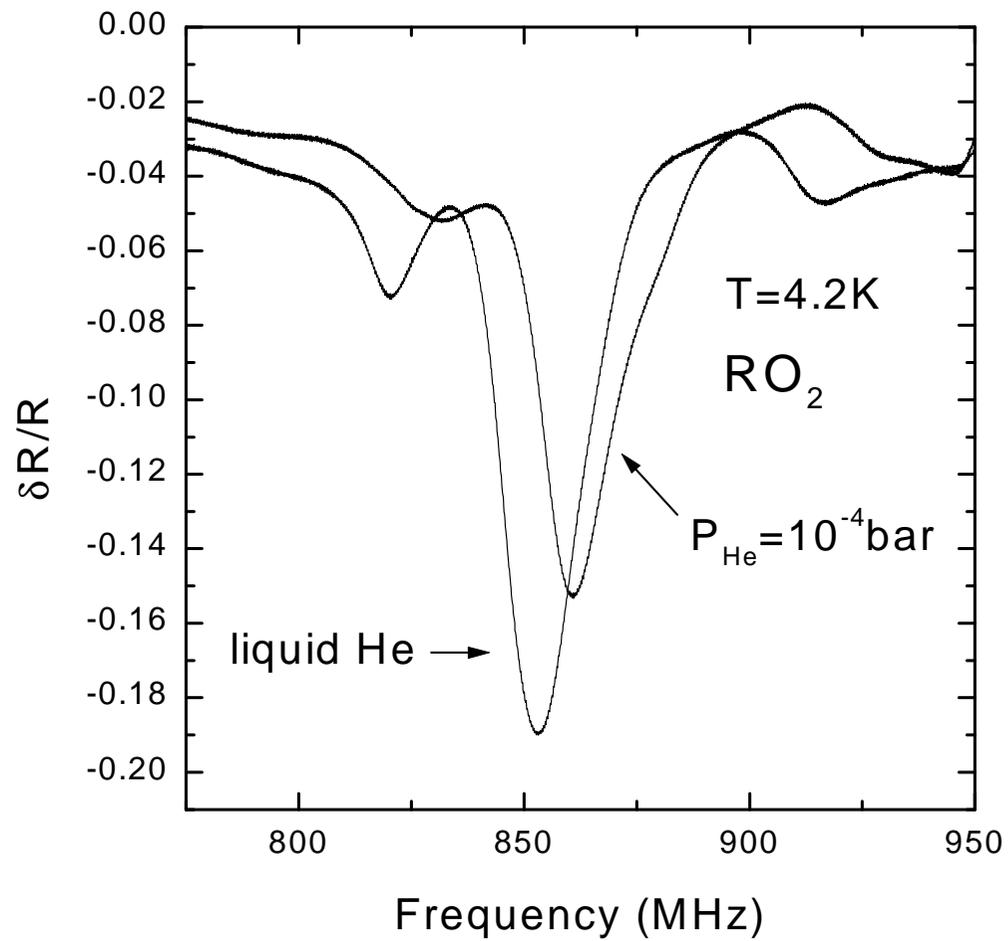
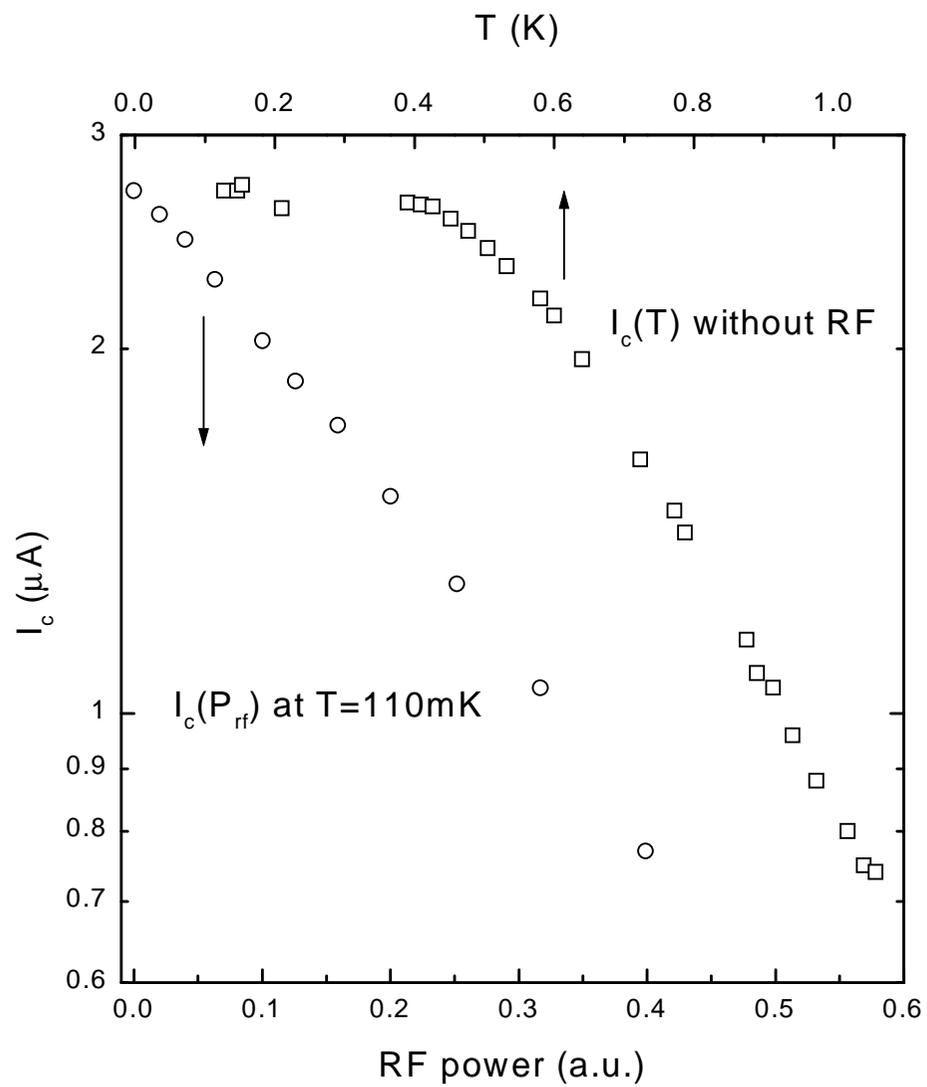
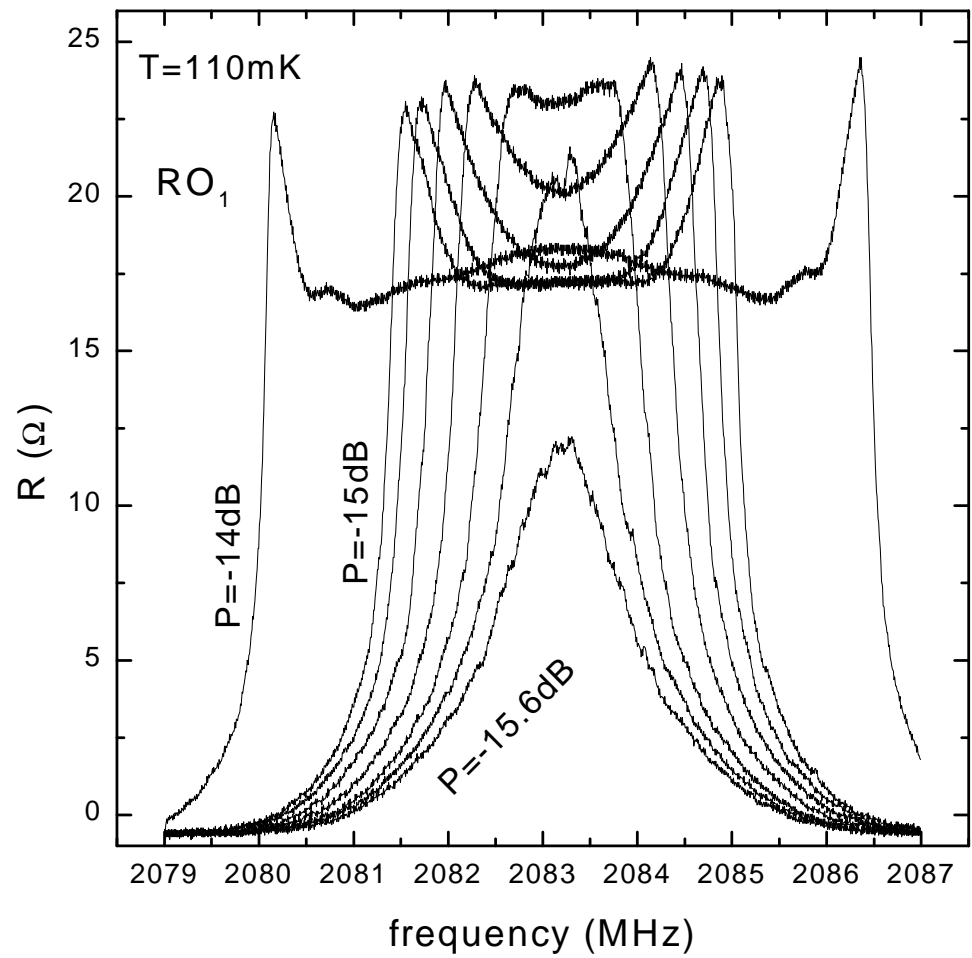
\end{document}